\documentclass{aa}
\idline{1}{1}

\usepackage{graphicx}
\usepackage {amssymb}
\usepackage {mathrsfs}
\usepackage {txfonts}

\begin{document}

\title {Evolution of the Barium abundance in the early Galaxy from a NLTE analysis of the Ba lines in a homogeneous sample of EMP stars.
\thanks{Based on observations obtained with the ESO Very Large
Telescope at Paranal Observatory (Large Programme ``First
Stars'', ID 165.N-0276; P.I.: R. Cayrel.}
}

\author {
S.M. Andrievsky\inst{1,2}\and
M. Spite\inst{1}\and 
S.A. Korotin \inst{2}\and 
F. Spite\inst{1}\and
P. Fran\c cois\inst{1}\and
P. Bonifacio\inst{1,3,4}\and
R. Cayrel\inst{1}\and
V. Hill\inst{1}
}

\institute{
GEPI, Observatoire de Paris,CNRS, Universit\'e Paris Diderot;  F-92125 Meudon Cedex, France,
 e-mail :  {\tt monique.spite@obspm.fr}
\and
Department of Astronomy and Astronomical Observatory, Odessa National University, Isaac Newton Institute of Chile, Odessa branch, Shevchenko Park, 65014 Odessa, Ukraine,
 e-mail : {\tt scan@deneb1.odessa.ua}
\and
CIFIST Marie Curie Excellence Team
\and
Istituto Nazionale di Astrofisica, Osservatorio Astronomico di Trieste, Via Tiepolo 11, I-34143 Trieste, Italy
}

\date {Received 1 September 2008; accepted 18 November 2008}
\titlerunning{NLTE Barium abundances in EMP stars}
\authorrunning{Andrievsky}

\abstract
{Barium is a key element in constraining  the evolution of the (not well understood) r-process in the
 first galactic stars  and currently the Ba abundances in these very metal-poor stars  were mostly measured under the Local Thermodynamical Equilibrium (LTE) assumption, which may lead in general   to an underestimation of Ba.}
{ We present here determinations of the barium abundance taking into account the non-LTE (NLTE) effects in a sample of extremely metal-poor stars (EMP stars): 6  turnoff stars and  35  giants.}
{ The NLTE profiles of the three unblended \ion{Ba}{ii} lines (4554\,\AA, 5853\,\AA, 6496\,\AA) have been computed. The  computations were made  with a modified version of the MULTI code, applied to an atomic model of the Ba atom with 31 levels of \ion{Ba}{i}, 101 levels of \ion{Ba}{ii}, and compared to the observations.
}
{The ratios of the NLTE abundances of barium relative to Fe are slightly shifted towards the solar ratio. In the plot of [Ba/Fe] versus [Fe/H], the slope of the regression line is slightly reduced as is the scatter. 
In the interval  $\rm -3.3 <[Fe/H] < -2.6$, [Ba/Fe] decreases with a slope of about 1.4 and a scatter close to 0.44. For $\rm [Fe/H] <-3.3$ the number of stars is not sufficient to decide whether  [Ba/Fe] keeps decreasing (and then CD-38:245 should be considered as a peculiar "barium-rich star") or if a plateau is reached as soon as $\rm [Ba/Fe] \approx -1$.
In both cases the scatter remains quite large, larger than what can be accounted for by the measurement and determination errors, suggesting the influence of a complex process of Ba production, and/or inefficient mixing in the early Galaxy.
}
{}
\maketitle
%
\section{Introduction} 

In the framework of the ESO "large programme First stars", very high quality spectra of extremely metal-poor (EMP) giants and turnoff stars were obtained with the high resolution spectrograph UVES fed with the VLT.
In this sample, 33 giants and 18 turnoff stars are "normal metal-poor stars" (not carbon-rich). Among them 22 giants and 10 turnoff stars have  $\rm [Fe/H]\leq -3$. The main parameters of these stars can be found in Cayrel et al. (\cite{CDS04}:"First stars V") and Bonifacio et al. (\cite{BMS07}: ``First stars VII'', \cite{BSC08}: ``First stars XII'').
In Fran\c cois et al. (\cite{FDH07}, "First stars VIII") the abundance of the neutron capture elements  has been studied in the sample of EMP giants. 
From an LTE analysis, it has been found that, generally speaking, the abundance ratios of these elements relative to iron ([X/Fe]) are very scattered, with sometimes a factor of more than one hundred between the abundance ratios of two stars with equal [Fe/H].
Moreover, on average, [Ba/Fe] increases from  about --1.5 to +0.2 while [Fe/H] increases from --3.6 to --2.6.
The important scatter of [Ba/Fe] (and [Sr/Fe]) vs. [Fe/H] has been confirmed by Bonifacio et al. (\cite{BSC08}, ``First stars XII'') where the abundances in the sample of giants are compared to the abundances in a sample of turnoff stars obtained in the same conditions.
On the other hand, from a similar LTE analysis, e.g. Burris et al. (\cite{BPA00}), Honda et al. (\cite{HAK04}) or Lai et al. (\cite{LBJ08}) found abundance ratios that,  in the same range of metallicity, show a similar behaviour. This general trend can also be traced from the rather low S/N measurements of Barklem et al. (\cite{BCB05}) in a very large sample of 373 metal-poor stars, albeit with a larger scatter.
 
Generally, the barium abundance in metal-poor stars has been investigated under the LTE approximation. However, the NLTE corrections can be very significant, in particular at low metallicity  (see Asplund, \cite{Asp-AR05}; Short \& Hauschildt, \cite{HS06}) they could be at least partly responsible for the large scatter of the abundance ratios of the heavy elements at low metallicity. In order to determine reliable trends and scatters, and to  properly trace the history of nucleosynthesis processes in the first stars, it is highly desirable to perform NLTE analyses.

There are several studies devoted to the determination of the NLTE barium abundance in metal-poor stars. Gigas (1988) performed NLTE modeling of the barium lines in the spectrum of the mildly metal deficient dwarf star Vega. Later, Mashonkina \& Bikmaev (1996), and Mashonkina et al. (1999) estimated abundance correction for the cooler and more metal deficient dwarfs (down to [Fe/H] = --2.5). This work was continued  by Mashonkina \& Gehren (2000), Mashonkina et al. (2003) and more recently by Mashonkina et al. (\cite{MZG08}).
In this last paper the authors study the abundance of the heavy elements in four very metal-poor stars with $\rm[Fe/H] \approx -2.5$: one giant, HD~122563, and three turnoff stars.
 Moreover Short \& Hauschildt (\cite{HS06}), performed a theoretical analysis of the  non-LTE effects on the Ba lines ($\lambda =$ ~4554, 5853, 6141, 6496 \AA) for metallicities between $\rm[Fe/H]=-1$ and --5. They used NLTE models for these computations, but an interesting point is that they could show that the NLTE profiles based on NLTE models differ negligibly from the NLTE profiles computed from LTE models (as is generally done).

We present here the results of a direct application of an NLTE analysis to the sample of extremely metal-poor stars (turnoff stars and giants) previously analyzed in the "First stars" programme. In our previous papers (Andrievsky et al. 2007, 2008) we reported 
the sodium and aluminium NLTE abundances in these same stars.

This sample covers the region of metallicities from $\rm [Fe/H] \approx -2.5 ~to -4.0$. Such metal-poor stars have not been investigated up to now with the aim of deriving their NLTE barium abundances.

\section{Observations and reduction.} 
The sample of stars and the observational data are the same as dicussed in previous papers :  First Stars V,  VII, VIII and XII (Cayrel et al., \cite{CDS04}; Bonifacio et al., \cite{BMS07}; 
Fran\c cois et al., \cite{FDH07}; and Bonifacio et al., \cite{BSC08}). In brief, the observations were performed with the VLT-UT2 and UVES (Dekker et al., \cite{DDK00}) with a resolving power $\rm R \approx 45000$. 
The signal-to-noise ratio in the spectra of the giants is very high:  $\sim 130$~per pixel (with an average of 5 pixels per resolution element) but it is lower for the turnoff stars (generally fainter), in particular in the $\lambda 4554\AA$ region where  the strongest Ba line is located, at the very end of the spectra obtained with the ``blue'' camera.

The spectra were reduced using the UVES context (Ballester et al. \cite{BMB00}) within MIDAS.

\begin {table*}[t]
\caption {Adopted model and NLTE barium abundance for our sample of stars. An asterisk after the name of the star means that the star is carbon-rich. In the last column is given the number of Ba lines used for the computation; the letter m indicates that the giant has been found "mixed" (see text).
}
\label {tabstars}
\begin {center}
\begin{tabular}{l l r r c r r r r r r}
~ & star & $\rm T_{eff},$(K)& log g& $\rm v_{t}, (km s^{-1})$ &  [Fe/H] & $\rm \epsilon (Ba)_{LTE}$&
$\rm \epsilon (Ba)_{NLTE}$&$\rm [Ba/H]_{NLTE}$& $\rm [Ba/Fe]_{NLTE}$& Rem\\

\hline
\\
& turnoff stars\\
\hline
   & BS 17570-063  & 6240&  4.8&  0.5&  -2.92&  -1.05&  -0.83&  -3.00&  -0.08&  1\\
   & CS 22948-093  & 6360&  4.3&  1.2&  -3.30&  -0.90&  -0.83&  -3.00&  +0.30&  1\\
   & CS 22966-011  & 6200&  4.8&  1.1&  -3.10&  -0.95&  -0.73&  -2.90&  +0.20&  1\\
   & CS 29506-007  & 6270&  4.0&  1.7&  -2.91&  -0.55&  -0.23&  -2.40&  +0.51&  1\\
   & CS 29506-090  & 6300&  4.3&  1.4&  -2.83&  -1.00&  -0.63&  -2.80&  +0.03&  1\\
   & CS 30301-024  & 6330&  4.0&  1.6&  -2.75&  -0.85&  -0.53&  -2.70&  +0.05&  1\\
\hline
\\
& giants\\
\hline
1  & HD 2796       & 4950&  1.5&  2.1&  -2.47&  -0.48&  -0.46&  -2.63&  -0.12&  3, m\\
2  & HD 122563     & 4600&  1.1&  2.0&  -2.82&  -1.59&  -1.45&  -3.62&  -0.80&  3, m\\
3  & HD 186478     & 4700&  1.3&  2.0&  -2.59&  -0.50&  -0.45&  -2.62&  -0.03&  3, m\\
4  & BD +17:3248   & 5250&  1.4&  1.5&  -2.07&  +0.75&  +0.48&  -1.69&  +0.38&  3, m\\
5  & BD -18:5550   & 4750&  1.4&  1.8&  -3.06&  -1.67&  -1.45&  -3.62&  -0.56&  3   \\
6  & CD -38:245    & 4800&  1.5&  2.2&  -4.19&  -2.82&  -2.50&  -4.67&  -0.48&  1, m\\
7  & BS 16467-062  & 5200&  2.5&  1.6&  -3.77&  $<-2.76$&  $<-2.35$&  $<-4.52$&  $<-0.75$&   \\
8  & BS 16477-003  & 4900&  1.7&  1.8&  -3.36&  -1.68&  -1.40&  -3.57&  -0.21&  2   \\
9  & BS 17569-049  & 4700&  1.2&  1.9&  -2.88&  -0.55&  -0.50&  -2.67&  +0.21&  3, m\\
10 & CS 22169-035  & 4700&  1.2&  2.2&  -3.04&  -2.10&  -1.95&  -4.12&  -1.08&  2, m\\
11 & CS 22172-002  & 4800&  1.3&  2.2&  -3.86&  -2.90&  -2.60&  -4.77&  -0.91&  1,  \\
12 & CS 22186-025  & 4900&  1.5&  2.0&  -3.00&  -0.85&  -0.72&  -2.89&  +0.11&  3, m\\
13 & CS 22189-009  & 4900&  1.7&  1.9&  -3.49&  -2.65&  -2.40&  -4.57&  -1.08&  1,  \\
14 & CS 22873-055  & 4550&  0.7&  2.2&  -2.99&  -1.31&  -1.15&  -3.32&  -0.33&  3, m\\
15 & CS 22873-166  & 4550&  0.9&  2.1&  -2.97&  -1.54&  -1.32&  -3.49&  -0.52&  3, m\\
16 & CS 22878-101  & 4800&  1.3&  2.0&  -3.25&  -1.70&  -1.35&  -3.52&  -0.27&  2, m\\
17 & CS 22885-096  & 5050&  2.6&  1.8&  -3.78&  -2.75&  -2.50&  -4.67&  -0.89&  1   \\
18 & CS 22891-209  & 4700&  1.0&  2.1&  -3.29&  -1.71&  -1.43&  -3.60&  -0.31&  3, m\\
19 & CS 22892-052* & 4850&  1.6&  1.9&  -3.03&  +0.11&  -0.03&  -2.20&  +0.83&  3   \\
20 & CS 22896-154  & 5250&  2.7&  1.2&  -2.69&  -0.05&  -0.13&  -2.30&  +0.39&  3   \\
21 & CS 22897-008  & 4900&  1.7&  2.0&  -3.41&  -2.28&  -2.15&  -4.32&  -0.91&  1   \\
22 & CS 22948-066  & 5100&  1.8&  2.0&  -3.14&  -1.95&  -1.75&  -3.92&  -0.78&  1, m\\
23 & CS 22949-037* & 4900&  1.5&  1.8&  -3.97&  -2.42&  -2.25&  -4.42&  -0.45&  1, m\\
24 & CS 22952-015  & 4800&  1.3&  2.1&  -3.43&  -2.63&  -2.35&  -4.52&  -1.09&  1, m\\
25 & CS 22953-003  & 5100&  2.3&  1.7&  -2.84&  -0.22&  -0.23&  -2.40&  +0.44&  3   \\
26 & CS 22956-050  & 4900&  1.7&  1.8&  -3.33&  -1.98&  -1.73&  -3.90&  -0.57&  1   \\
27 & CS 22966-057  & 5300&  2.2&  1.4&  -2.62&  -0.73&  -0.73&  -2.90&  -0.28&  2   \\
28 & CS 22968-014  & 4850&  1.7&  1.9&  -3.56&  $<-3.20$&  $<-2.80$&  $<-4.97$&  $<-1.41$&    \\
29 & CS 29491-053  & 4700&  1.3&  2.0&  -3.04&  -1.80&  -1.52&  -3.69&  -0.65&  3, m\\
30 & CS 29495-041  & 4800&  1.5&  1.8&  -2.82&  -1.34&  -1.06&  -3.23&  -0.41&  3   \\
31 & CS 29502-042  & 5100&  2.5&  1.5&  -3.19&  -2.75&  -2.40&  -4.57&  -1.38&  1   \\
32 & CS 29516-024  & 4650&  1.2&  1.7&  -3.06&  -1.83&  -1.45&  -3.62&  -0.56&  3   \\
33 & CS 29518-051  & 5200&  2.6&  1.4&  -2.69&  -1.01&  -0.60&  -2.77&  -0.08&  3, m\\
34 & CS 30325-094  & 4950&  2.0&  1.5&  -3.30&  $<-3.05$&  $<-2.85$&  $<-5.02$&  $<-1.72$&    \\
35 & CS 31082-001  & 4825&  1.5&  1.8&  -2.91&  +0.39&  +0.05&  -2.12&  +0.79&  3   \\
\hline   
\end {tabular}  
\end {center}  
\end {table*}

Table 1 gives the NLTE barium abundances for our sample. They could be determined for only  six (out of 18) turnoff stars, because even their strongest lines of Ba are very weak.
The $\rm \epsilon (Ba)_{LTE}$ value, given in column 6 for comparison,  has been borrowed from Fran\c cois et al (2007) for the giants and Bonifacio et al. (2008) for the turnoff stars.
In the last column is given the total number of Ba lines used for the computation. If n=3, all the three lines 4554, 5853, and 6496\AA~ have been computed, if n=2 only 4554 and 6496\AA~ could be used and if n=1 only 4554\AA. For the giants, an "m" indicates that the star has been found "mixed" in Spite et al. (\cite{SCP05}). In this table, for the determination of the relative LTE and NLTE values we have adopted  for the solar values $\rm log \epsilon(Ba) = 2.17$ (Asplund et al. \cite{ASP05}) at variance with Fran\c cois et al. \cite{FDH07} who adopted $\rm log \epsilon(Ba) = 2.13$ (Grevesse \& Sauval \cite{GS00}).

\section{NLTE calculations.}

The NLTE profiles of the barium lines were computed using a 
modified version of the MULTI code (Carlsson \cite{Carl86}). 
The modifications are described in Korotin et al. (\cite{Kor99}).

For these computations we used Kurucz's 1992 models (ATLAS9), after checking on some typical stars that the use of MARCS models (Gustafsson et al. \cite{GEE08}) as in Cayrel et al. (\cite{CDS04}) would not introduce any significant difference.

As a rule, in the turnoff stars with $\rm[Fe/H]<-2$, the barium abundance can be determined only from the resonance line at 4554 \AA, but even this line becomes too weak at $\rm [Fe/H] <-3.0$. In the giants, three lines can be traced down to $\rm [Fe/H] \approx -3.0$, the resonance line and the two subordinate lines at 5853 and 6496~\AA.; at still lower metallicity only two lines (4554 and 6496 \AA~) are still detectable. 

\subsection{Atomic model of Ba}
\begin {figure}[ht]
\begin {center}
\resizebox  {8.0cm}{7.0cm}
{\includegraphics{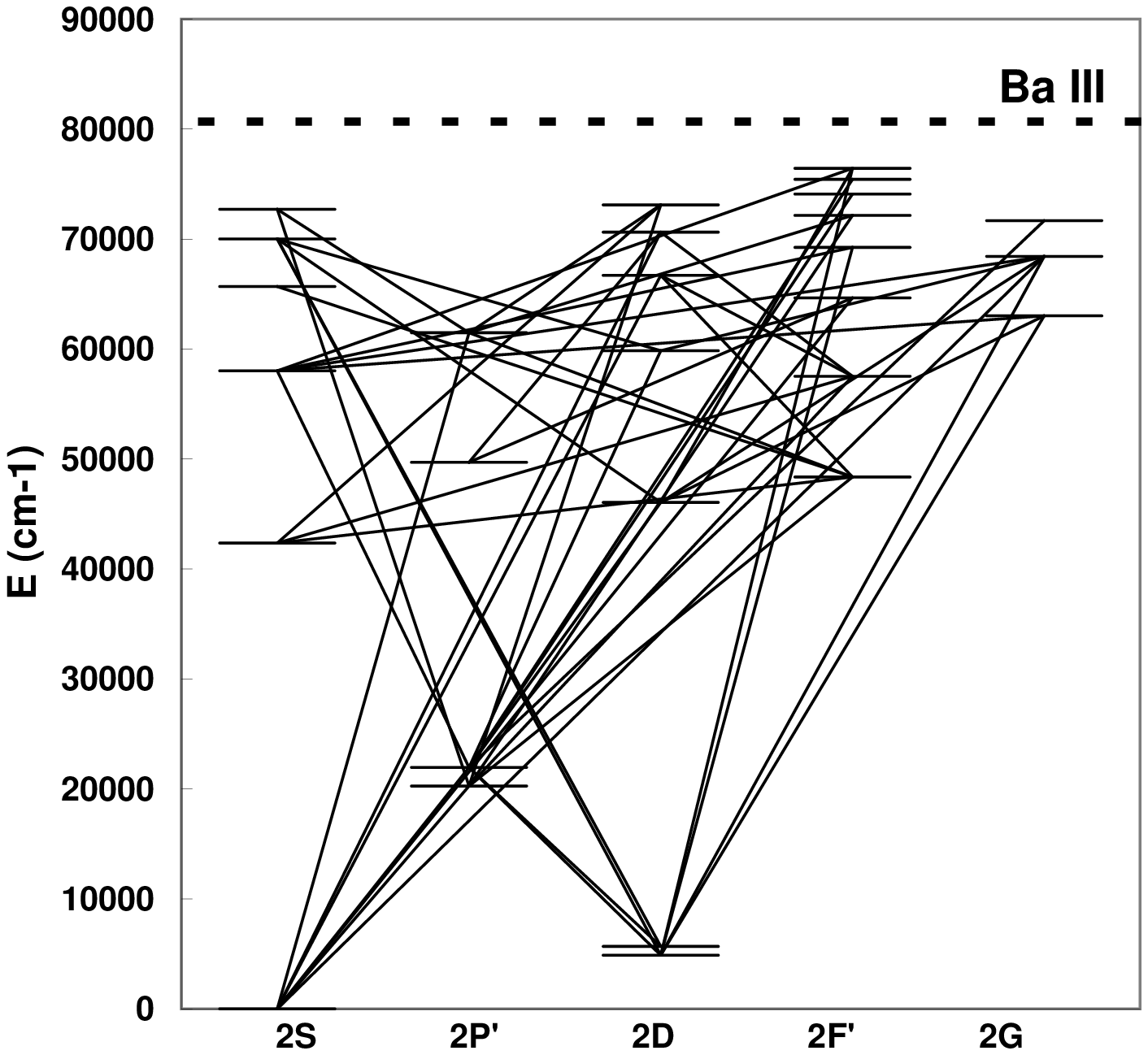}}
\caption{The Grotrian diagramme for \ion{Ba}{ii}, only those
transitions that were considered in detail, are indicated.}
\label {grot}
\end {center}
\end {figure}

Our barium model contains 31 levels of \ion{Ba}{i},  101
levels of \ion{Ba}{ii} with $n < 50$, and the ground level of \ion{Ba}{iii} ion. 
In the detailed consideration we included 91 bound-bound transitions between the first 28 levels of \ion{Ba}{ii} with $n < 12$ and $l < 5$. 
The other levels were used for the particle number conservation. The fine structure  of the levels $\rm 5d^2D$ and $\rm 6p^2P^0$ was taken into account;  the other levels were treated as single.
The corresponding Grotrian diagram is shown in Fig.\ref{grot}. Only those transitions that were considered in detail, are indicated. The energies of the levels are from Curry (\cite{Curry04}). The oscillator strengths of the bound-bound transitions are from Wiese \& Martin(\cite{WM80}), Warner (\cite{War68}), and Miles \& Wiese (\cite{MW69}). For the transitions between the seven low-lying levels we used the data from Davidson et al. (\cite{Dav92}).

\begin {figure}[ht]
\begin {center}
\resizebox  {8.5cm}{3.5cm}
{\includegraphics {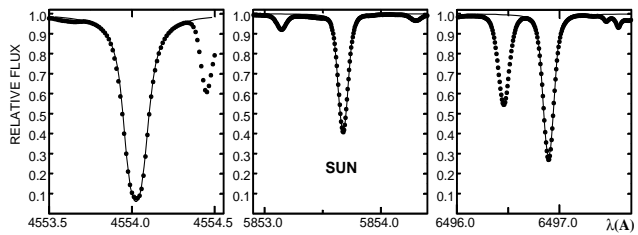} }
\caption{Profile fitting for the barium lines in the solar spectrum
with $\rm log \epsilon(Ba) = 2.17$.}
\label {spesol}
\end {center}
\end {figure}

\begin {figure}[ht]
\begin {center}
\resizebox{\hsize}{!}
{\includegraphics {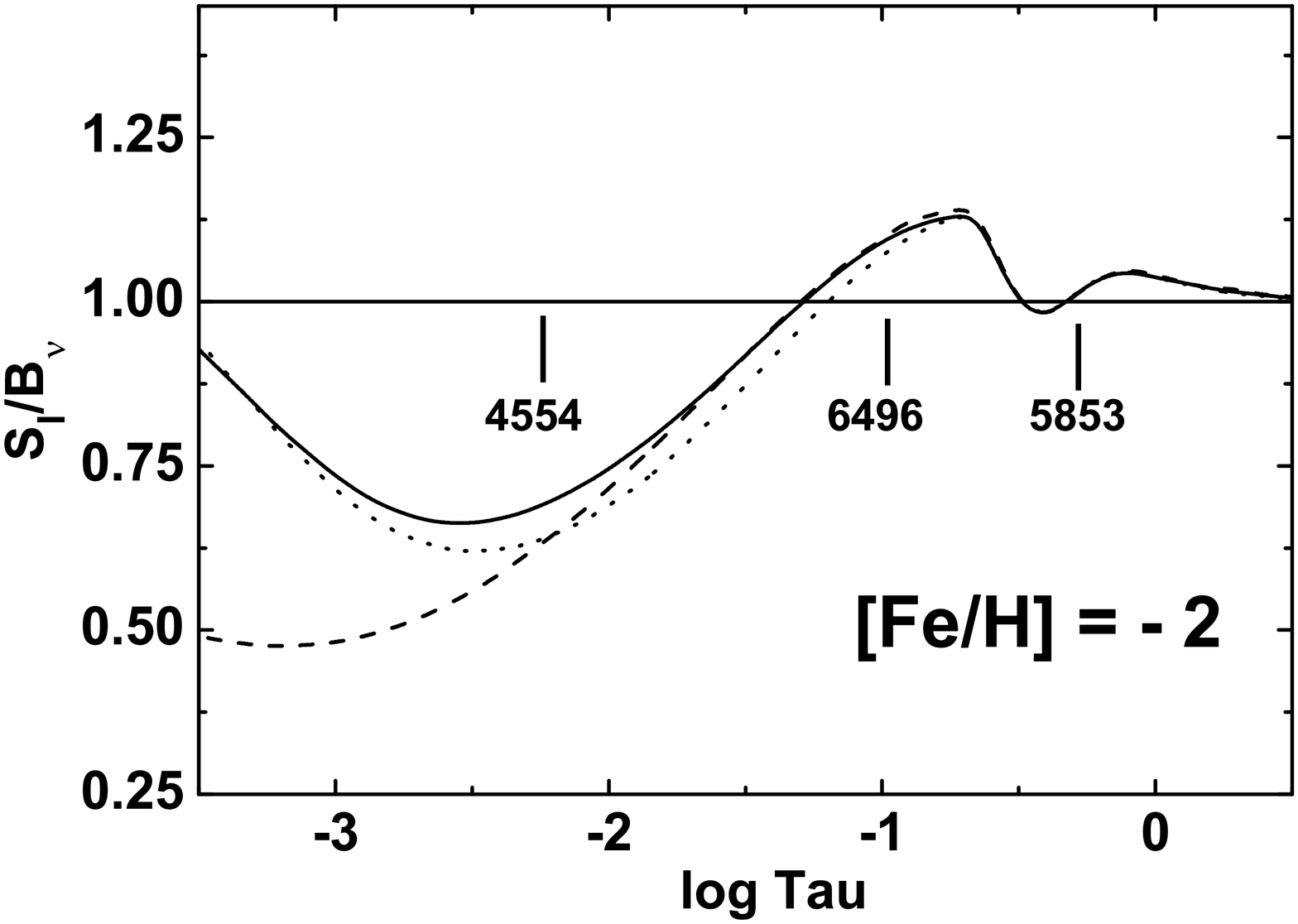}}
\resizebox{\hsize}{!}
{\includegraphics {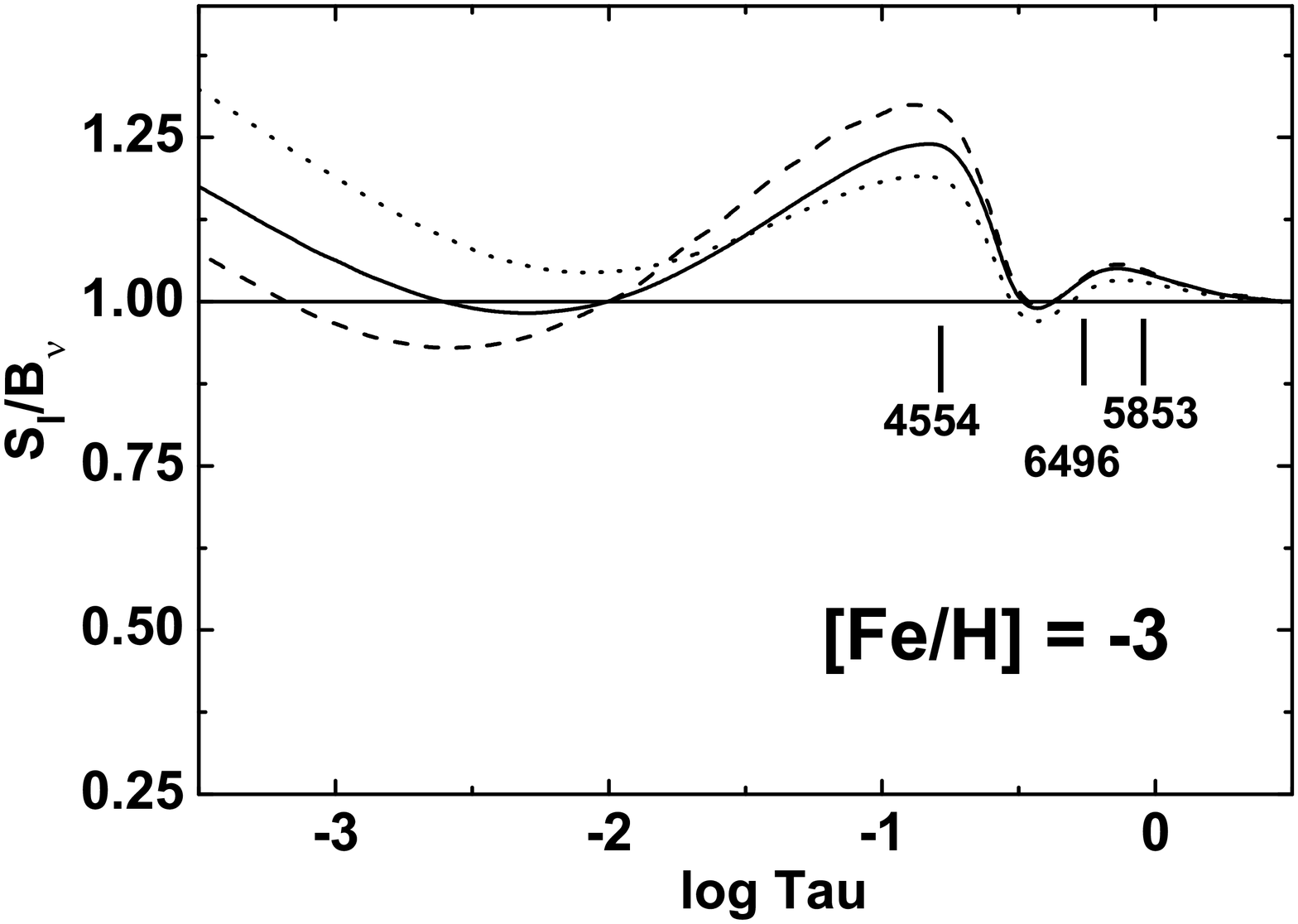}}
\caption{Ratio $\rm S_{\it{l}}/B_{\nu}$ (Source function over Planck
function) vs.  the optical depth for the models $\rm T_{eff}=5250K$,
$\rm \log~g=1.5$, $\rm [Fe/H]=-2.0$ and $\rm -3.0$.  The dashed line
represents the 4554~\AA~line, the solid line the 5853~\AA~ line and
the dotted line the 6496~\AA~line.
}
\label {sb}
\end {center}
\end {figure}

\begin {figure}[ht]
\begin {center}
\resizebox  {8.5cm}{4.5cm} 
{\includegraphics {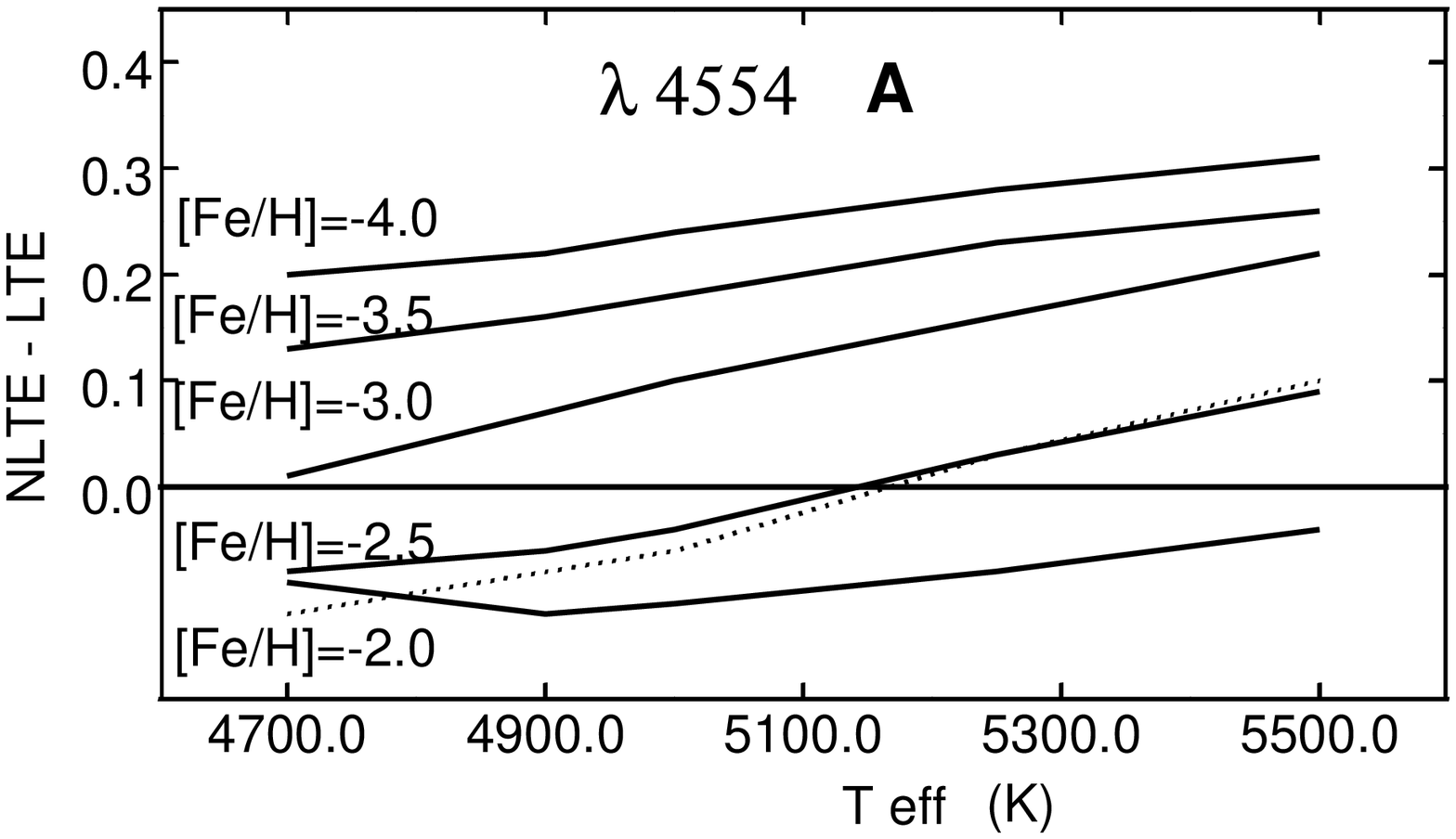}}
\resizebox  {8.5cm}{4.5cm}
{\includegraphics {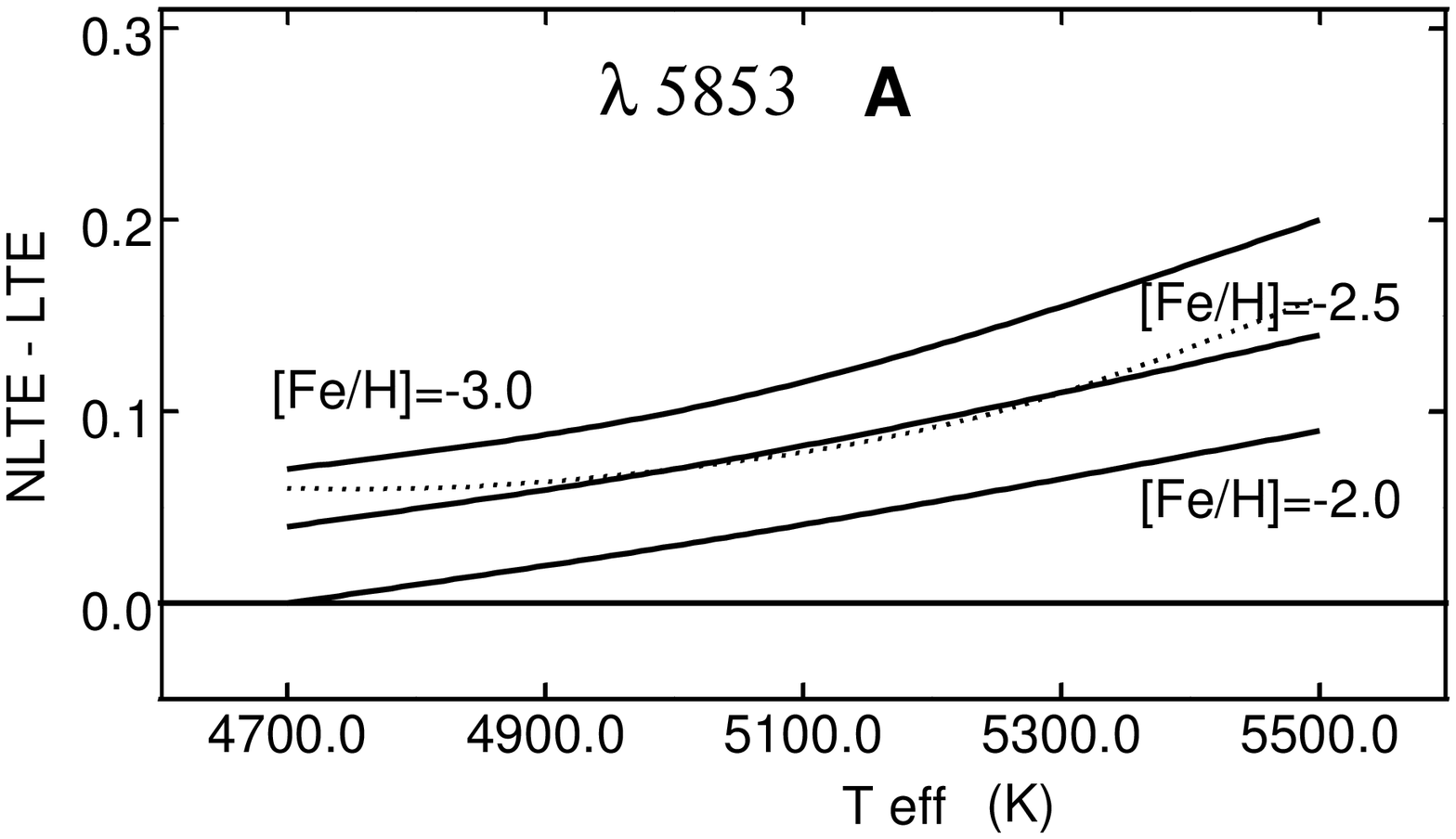}}
\resizebox  {8.5cm}{4.5cm}
{\includegraphics {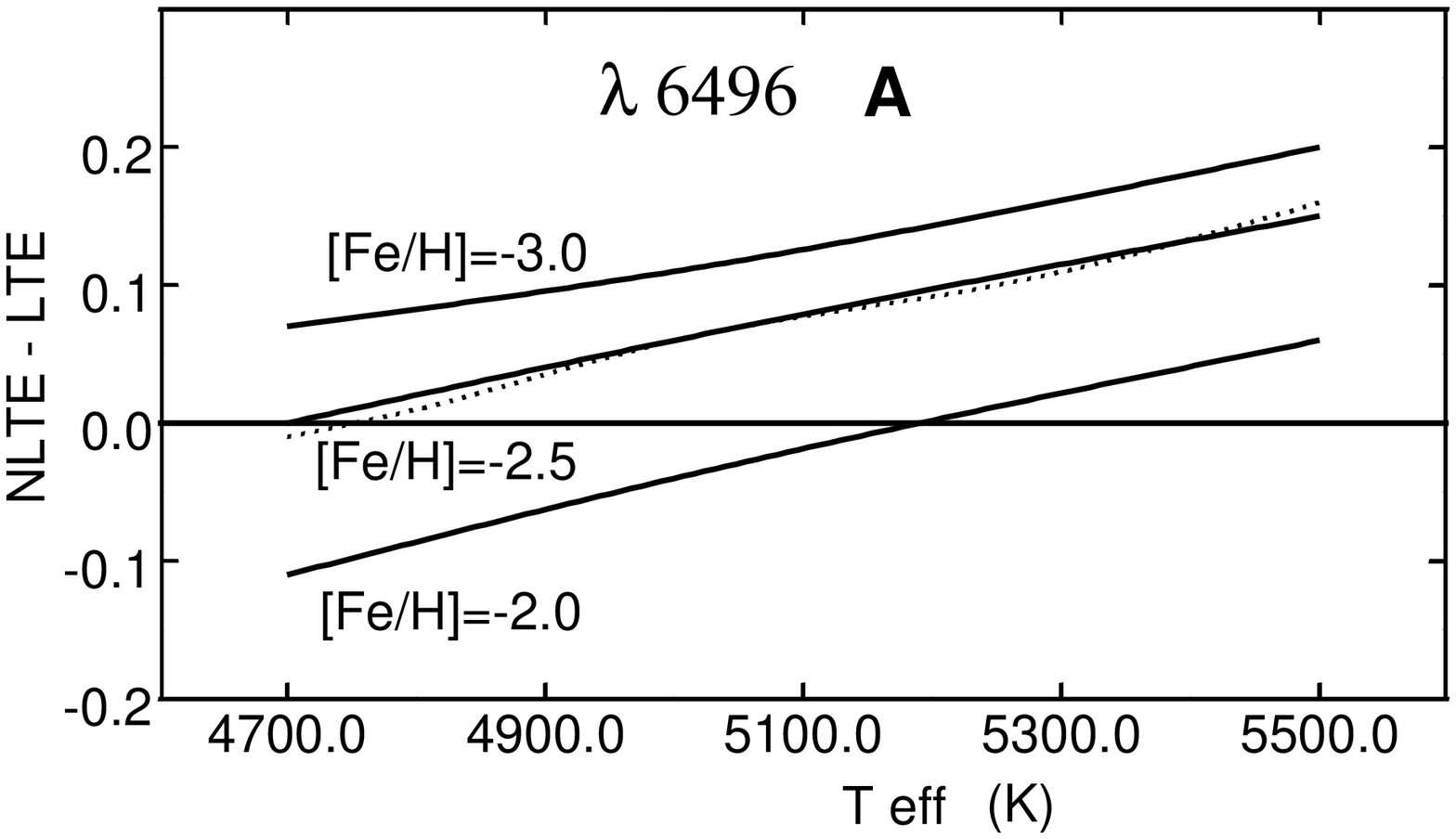}}
\caption{NLTE corrections computed as a function of the temperature of
the model, for several metallicities.  The calculations are performed
for $\log~g = 1.5$ and [Ba/Fe]=--0.5 (solid lines).  However, the
computations have been done also for [Fe/H]= --3.0 and [Ba/Fe] = 0.0,
thus [Ba/H] = --3.0 (dotted lines).  It can be seen that the
correction is then very similar to the correction computed for [Fe/H]
= --2.5 and [Ba/Fe] = --0.5 (thus also with [Ba/H]= --3.0).  The main
parameter that determines the NLTE correction, is indeed [Ba/H] and
{\it not} [Fe/H].
}
\label{nlte-cor}
\end {center}
\end {figure}

\begin {figure}[ht]
\begin {center}
\resizebox  {8.5cm}{4.5cm} 
{\includegraphics {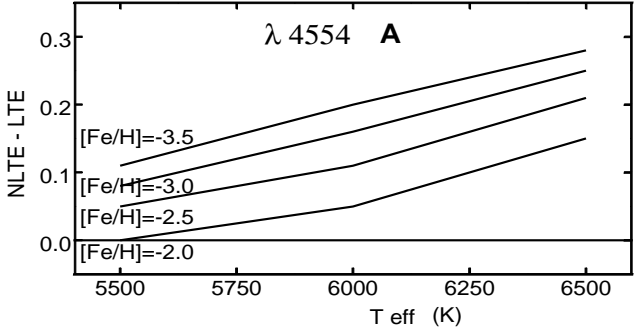}}
\caption{NLTE corrections computed as a function of the temperature of
the dwarf models, for several metallicities.  All calculations are
performed with [Ba/Fe]=0.0 and $\log~g = 4.5$.}
\label {nlte-cordw}
\end {center}
\end {figure}

\begin {figure}[ht]
\begin {center}
\resizebox  {8.5cm}{4.5cm} 
{\includegraphics {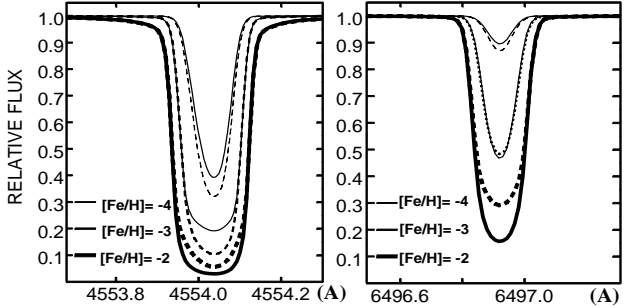} }
\caption{LTE (dashed lines) and NLTE (solid lines) profiles calculated
for T$_{\rm eff} = 4800$\,K, [Ba/Fe]=0.0 and $log~g = 1.5$.}
\label {nlte-pr}
\end {center}
\end {figure}

\begin {table}[t]
\caption {Parameters of the barium lines.  $C_{6}= \Delta \omega ~r^6 / 2 \pi$,  following (Uns\"old, \cite {Uns68}), in $\rm cm^6 ~s^{-1}$.
}
\label {hfs}
\begin {center}
\begin{tabular}{lc@{~~}c@{~~~}r@{~~~~} c@{~~}c@{~~~}c@{~~~}c@{ }c@{ }r@{ }r@{
}c@{ }c@{ }c@{ }c}
\hline
$\rm \lambda (\AA)$& $E_{i}$& $E_{k}$& HFS           & \it f & \it f& $\rm \gamma_{rad}$&$\rm log C_{6}$\\
                   &       &         &$\Delta\lambda$&  solar& 50:50& $s^{-1}$&  \\
                   &  eV   &   eV    &   m\AA        \\
\hline
4554.03& 0.00& 2.72&  0&  0.597&  0.364&  $1.58~10^8$&  -31.3\\
       &     &     & 18&  0.081&  0.227& \\
       &     &     &-34&  0.049&  0.137& \\
\\
5853.70& 0.60& 2.72& - &  0.025&  -    &  $1.58~10^8$&  -30.4\\
\\
6496.90& 0.60& 2.51&  0&  0.086&  -    &  $1.25~10^8$&  -31.1\\
       &     &     & -4&  0.012&  -    &\\
       &     &     &  9&  0.007&  -    &\\ 
\hline  
\end {tabular}  
\end {center}  
\end {table}

\subsection{The different constants}

A cause of uncertainty in the NLTE analysis of the barium spectrum is 
the scarce information about the photoionization 
cross-sections for the different levels. For the majority of the levels 
we used the results of the scaled Thomas-Fermi method application 
(Hofsaess \cite{Hof79}). For g-levels and levels with n = 11 we used the 
hydrogen-like approximation  (Lang, \cite{Lang98}, eq. 1.230). 

Effective excitation electron collisional strengths for the
transitions between the first levels ($\rm 6s^2S$, $\rm 5d^2D$ and
$\rm 6p^2P^0$) were used following Schoening \& Butler
(\cite{SBut98}).  Experimental cross-sections for the transitions
$\rm6s^2S-7s^2S$ and $\rm6s^2S-6d^2D$ were taken from Crandall et al.
(\cite{Cr74}).  Collisional rates for the transitions between
sublevels $\rm 5d^2D$, $\rm 6p^2P^0$ and $\rm7s^2S$, $\rm 6d^2D$, and
between $\rm7s^2S$ and $\rm6d^2D$ were estimated with the help of the
corresponding formula from Sobelman et al.  (\cite{Sob81}).  For the
rest of the allowed transitions, we used the van Regemorter
(\cite{Reg62}) formula, while for the forbidden transitions the
Allen's (\cite{All73}) formula was used.

The rate of the collisional ionization from the ground level 
of \ion{Ba}{ii} was calculated using the corresponding formula in 
Sobelman et al. (\cite{Sob81}). The use of experimental data of Peart et al. 
(\cite{Peart89}) and Feeney et al. (\cite{Feen72})  give the same results. 
For the other levels, Drawin's (\cite{Draw61}) formula was used. 

Inelastic collisions of barium atoms with hydrogen atoms may play a
significant role in the cool star atmospheres.  They have been taken
into account through the formula of Steenbock \& Holweger
(\cite{StHol}).  A correcting factor of 0.1 was derived as a result of
the experimental fitting of the barium lines in solar spectrum.
Collisions with atomic hydrogen for the forbidden transitions were not
taken into account, but we checked that the resulting uncertainty
about the equivalent width of the lines, computed by the method of
Takeda (\cite{Taked91}), does not exceed 0.5\%.

The odd barium isotopes have hyper-fine splitting of their levels and
thus several HFS components for each line.  As was demonstrated by
Mashonkina et al.  (\cite{Mash99}), a three-component structure is
sufficient to describe the 4554~\AA~ barium line (see also Rutten
1978).  This line was fitted in the solar spectrum by adopting the
even-to-odd abundance ratio of 82:18 (Cameron \cite{Cam82}).  Since
metal deficient stars are supposed to include a larger fraction of the
material synthesized in SNe II through the pure r-process, we used an
even-to-odd ratio 50:50 for the stars.  Nevertheless, we checked that
the resulting difference in the equivalent widths of the 4554\AA~ line
due to this change in the even-to-odd ratio is very small.  (As a
consequence, from our spectra it is not possible to estimate the
abundance ratios of the different isotopes of Ba from our spectra,
higher resolution would be necessary.)

Radiative damping constants are from Mashonkina \& Bikmaev
(\cite{Mash96}), Stark broadening parameters are from the VALDatabase
(http://ams.astro.univie.ac.at/\~{ }vald).  Since the classical van
der Waals formula underestimates the effect of the interaction with
neutral particles, the broadening parameters were found by comparing
the observed lines in the solar spectrum (Kurucz et al.  \cite{Kur84})
with a synthetic spectrum computed with the solar model of Kurucz
(\cite{Kur96}).  The best agreement was obtained for $\rm log
\epsilon(Ba) = 2.17$ (Fig.  \ref{spesol}), in good agreement with the
estimation of the solar abundance of barium of Asplund et al.
(\cite{ASP05}).

The oscillator strengths of the lines components and the corresponding
HFS shifts, as well as the broadening parameters are given in Table
\ref{hfs}.  For the Van der Waals constant $C_{6}$ we have adopted the
definition of Uns\"old (\cite {Uns68}, eq.  82,47): $C_{6}= \Delta
\omega ~r^6 / 2 \pi$.

\subsection{Non-LTE effects}
The profiles of the barium lines are rather sensitive to the NLTE
effects, in particular in metal-poor stars with rather high
temperature.  The NLTE correction sign changes when the temperature
and the metallicity change.  Mashonkina et al.  (\cite{Mash99})
discuss in detail this behaviour in detail for turnoff metal-poor
stars.  Since the abundance of barium at a given metallicity is very
scattered, it is important to note that the NLTE corrections are very
sensitive not only to the metallicity of the model but also to the
barium abundance itself (the equivalent width of the line).

-- The resonance line or the low excitation subordinate lines that are
used for the determination of the barium abundance can show
substantial nonequilibrium excitation effects.  In a metal-poor
atmosphere, since there are few electrons and thus few collisions, the
radiative processes cause the departure from equilibrium in the atomic
level populations even in the deep atmospheric layers.  If the mean
intensity of the radiative field in the line frequencies J$_{\nu}$
exceeds the Planck intensity B$_{\nu}$, then an enhanced
photoexcitation depopulates the lower atomic levels and overpopulates
the upper ones (this ``pumping'' is described e.g. in Bruls et al.
\cite{Bruls92,} or Asplund \cite{Asp-AR05}).  The depth of the
effective formation of the radiation in b-b transitions changes
significantly with the metallicity (and the corresponding barium
abundance) of the star (Fig.  \ref{sb}).  This explains the dependence
of the NLTE correction on [Fe/H] (Fig.  \ref{nlte-cor} and
\ref{nlte-cordw}).  The lower the metallicity of the model, the deeper
in the atmosphere is the line formation level (Fig.  \ref{sb}).

-- Another mechanism causing the deviations from LTE is the so called
UV overionization from the 6p level.  In the atmosphere of turnoff or
RGB stars, Ba~II is the dominant species and generally this implies
that overionization does not significantly affect the line formation.
But for extremely metal-poor stars, already at $\tau_{5000} = 1$ the
mean intensity exceeds the Planck intensity at the frequency of the
ionization threshold of the 6p level.  As a result this level is
significantly depopulated.

These effects are very sensitive to the metallicity of the model, the
physical conditions and the characteristics of the radiation in the
layers where the barium lines are formed.  In Fig.  \ref{sb} the ratio
of the source function to the Planck function is plotted for two giant
stars with the same parameters T$_{\rm eff} = 5250$\,K, $\log~g =
1.5$, V$_{\rm t} = 2$~km~s$^{-1}$, [Ba/Fe] = --0.5, but with different
metallicities: [Fe/H] = --2.0 and --3.0.  The formation depth of the
three lines used for the determination of the Ba abundance ($\rm
\lambda = 4554, 5853, 6496\AA$) is indicated in the figure.  It can be
seen that $\rm S_{\it{l}}/B_{\nu}$ is larger than 1 at the depth of
the line formation for the three considered lines for [Fe/H]= --3.0.
In this case, NLTE effects make the line weaker.  The model with a
higher metal abundance produces another picture.  While two lines
5853~\AA~ and 6496~\AA~ are formed in that region where $\rm
S_{\it{l}}/B_{\nu}> 1$, the resonance line is formed significantly
higher in atmosphere, and for this line $\rm S_{\it{l}}/B_{\nu}< 1$.
This line is made stronger as a result of the departure from LTE.

As an illustration we show in Fig.  \ref{nlte-cor} (giants), and
\ref{nlte-cordw} (turnoff stars) how the NLTE correction depends on
the metallicity for different effective temperatures.  For the giant
stars we used [Ba/Fe]= --0.5, while for dwarfs [Ba/Fe] = 0.0.  These
values are typical of the observations.  (In turnoff stars if [Fe/H]=
--3.0 and [Ba/Fe]= --0.5, the strongest barium line is very weak ($\rm
W <4m\AA$) and since it is located at the end of the spectra it is
generally not detectable).  In fact the main parameter which
determines the NLTE correction is [Ba/H], as can be seen in Fig.
\ref{nlte-cor}.  Our corrections for the 4554~\AA~line are in good
agreement with those of Mashonkina \& Bikmaev (\cite{Mash96}) computed
for dwarfs and [Fe/H]= --2.

In giant stars (Fig.  \ref{nlte-cor}), the NLTE correction for the two
lines at 4554 and 5853~\AA~ can be positive or negative depending on
the effective temperature.  A similar conclusion was obtained by Short
\& Hauschildt (\cite{HS06}).  For the line 4554~\AA~ we found that LTE
and NLTE profiles coincide at [Fe/H]=--2.5, while Short \& Hauschildt
(\cite{HS06}) found such a coincidence at [Fe/H]= --3.0.

Our computations (NLTE line transfer with an LTE background model)
cannot be directly compared to those of Short \& Hauschildt who used
NLTE line transfer on an NLTE background model.  In spite of this, the
predictions of the two are qualitatively similar, with moderate
quantitative disagreement.  Part of the disagreement may, indeed, be
due to the difference in the adopted background models, but part of it
may also be due to the differences in the model atoms employed in the
two computations.  Some examples of the calculated profiles are given
in Fig.  \ref{nlte-pr}.

\subsection{Line profile fitting}
\begin {figure}[ht]
\begin {center}
\resizebox  {8.5cm}{6.cm} 
{\includegraphics {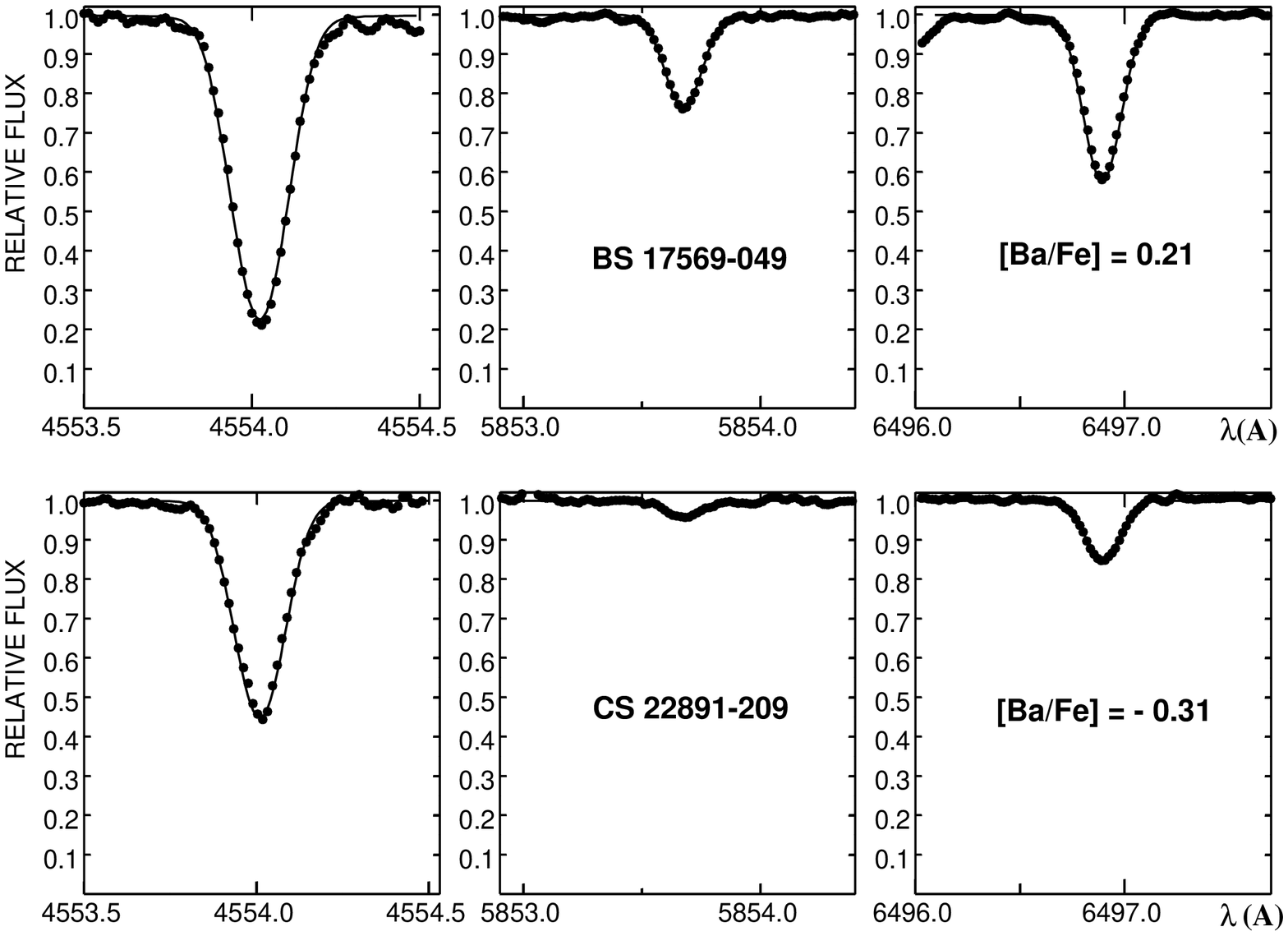} }
\resizebox  {4.0cm}{3.cm} 
{\includegraphics {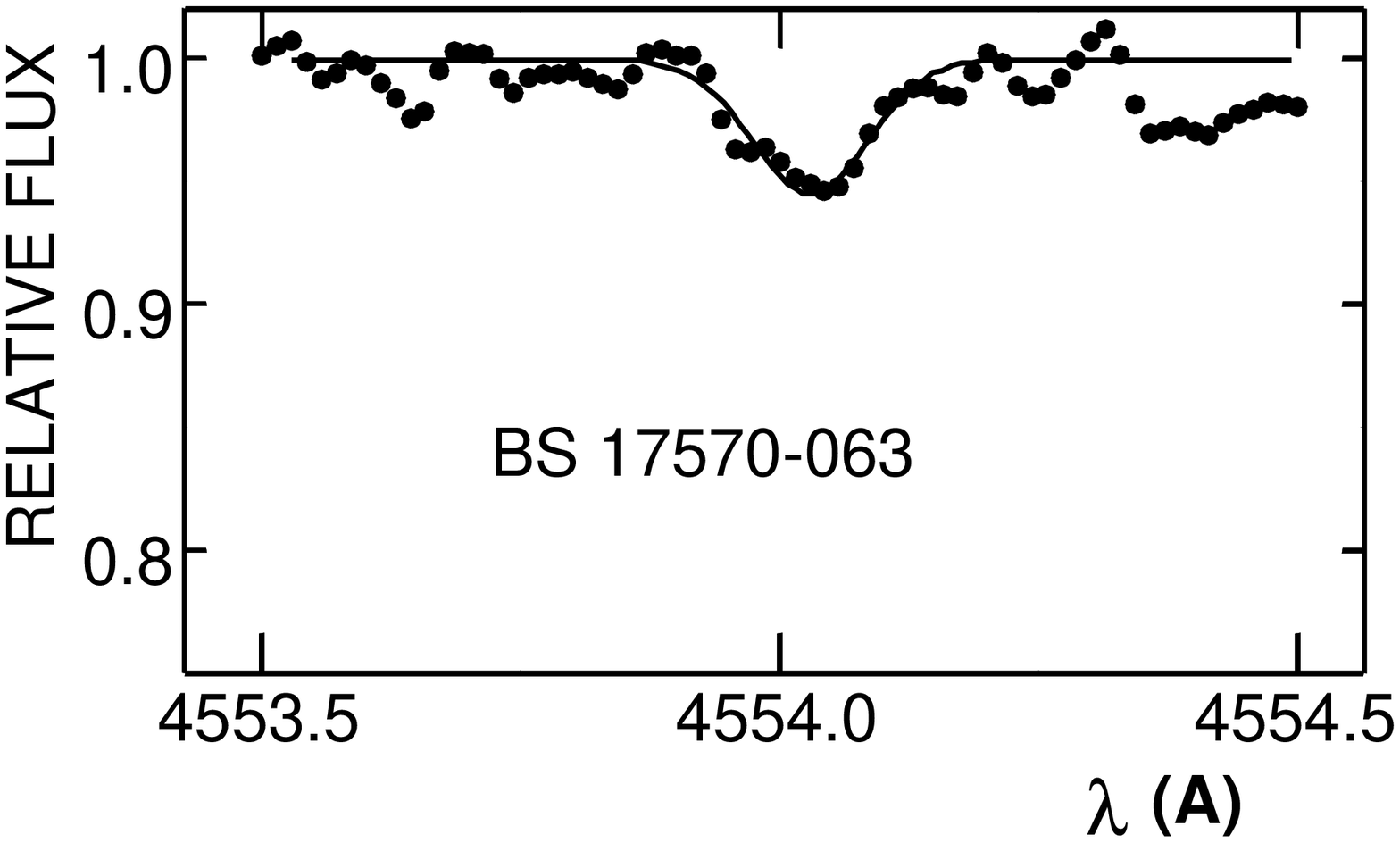} }
\caption{Profile fitting for the barium lines in two giant stars (BS~17569-049 
and CS~22891-209) and a turnoff star (BS~17570-063). In this last case only 
the 4554\AA~ line is visible.}
\label {spestars}
\end {center}
\end {figure}

It should be stressed that the NLTE corrections given in
Fig.~\ref{nlte-cor} or \ref{nlte-cordw} should not be used to
determine a precise value of the NLTE abundance of barium in
metal-poor stars, since this correction depends critically on several
parameters ([Ba/H], $\rm v_{t}$).  The best way to determine the Ba
abundance is to calculate the NLTE profiles directly, and to compare
them to the observed profiles.  Such an approach was used in this
paper.  The profile fitting for some stars is displayed in
Fig.~\ref{spestars}.

The mean NLTE barium abundances of the stars are listed in Table
\ref{tabstars} .  The small difference between the [Ba/Fe] ratio found
for HD~122563 in Mashonkina et al.  (\cite{MZG08}) and in our paper,
can be explained by the different values adopted for the gravity.
When several Ba lines are detected in the spectra of a star, the
abundances deduced from the different lines agree within $\rm \pm 0.1
dex$.  This value can be considered as the observational error.  The
total error is the quadratic sum of this error and of the error due to
the stellar parameter uncertainties.  The largest uncertainty arises
from the uncertainties in the temperature and the gravity of the
stars.  We estimate that the total error on [Ba/Fe] is close to
0.2~dex.

\begin {figure}[ht]
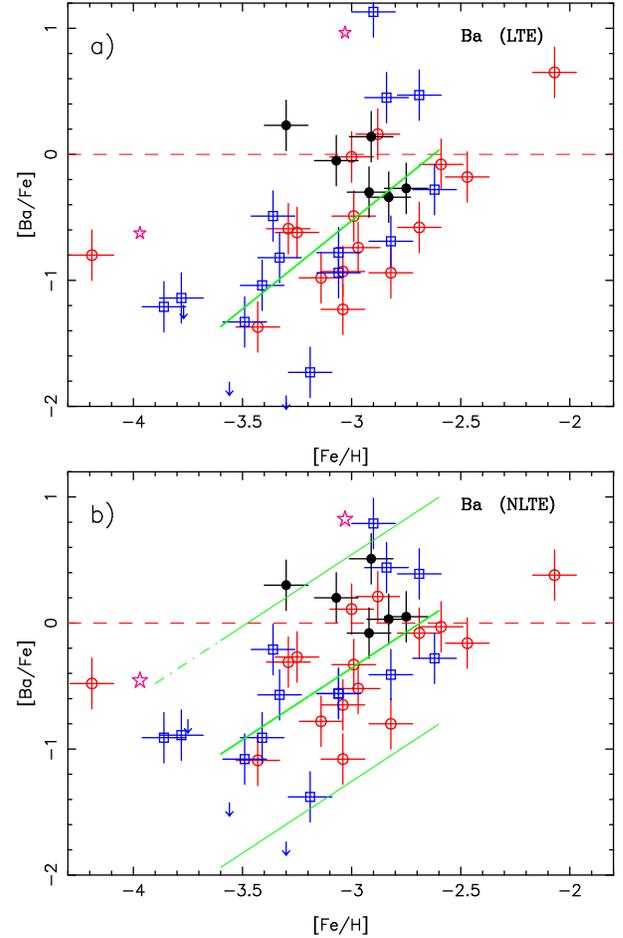

\begin {center}
\resizebox  {8.0cm}{6.2cm}  
{\includegraphics {0894fi8a.ps} }
\resizebox  {8.0cm}{6.2cm} 
{\includegraphics {0894fi8b.ps} }
\caption{[Ba/Fe] vs.  [Fe/H] in our sample of stars computed with the
LTE hypothesis (plot a) and taking into account the NLTE effects (plot
b).  The values of [Ba/Fe] have been computed with $\rm log
\epsilon(Ba) = 2.17$ (Asplund et al., \cite{ASP05}) at variance with
Fran\c cois et al.  (\cite{FDH07}) who adopted $\rm log \epsilon(Ba) =
2.13$.  The filled circles represent the turnoff stars, the open
circles (in red in the electronic version of the paper) the RGB mixed
giants and the open squares (in blue) the unmixed RGB giants.  The two
(``pink'') star-symbols represent the two carbon-rich stars we have
measured (see Table \ref{tabstars}).  The thick (green) lines
represent the mean slopes computed between [Fe/H]= --3.6 and [Fe/H]=
-- 2.6.  The slope of the regression line is slightly shallower when
non-LTE effects are taken into account.  In plot b) (NLTE) the thin
(green) lines are drawn at a distance of two sigmas from the
regression line.  The two carbon-rich stars are outside the two sigmas
limit.  If we consider only the ``normal'' stars (not carbon rich)
then 3 stars are outside this limit on the ``Ba-rich'' side.  }
\label{bafe}
\end {center} 
\end {figure}

\section{Discussion}

\subsection{Global trend}

The behavior of the ratio [Ba/Fe] vs.  metallicity for the
investigated stars is displayed in Fig.~\ref{bafe}.

The turnoff and the giant stars show a similar behaviour (as judged
from the few dwarfs for which Ba was measurable).  There is no
difference in the behaviour of mixed giants (stars located above the
Luminosity Function Bump : Spite et al., \cite{SCP05}, \cite{SCH06})
and the low RGB giants.  Barium is not affected by mixing along the
RGB.

In Fran\c cois et al.  (\cite{FDH07}) it has been shown from an LTE
analysis that in the interval $\rm - 3.5 <[Fe/H] < -2.6$, the
abundance of barium decreases rapidly with the metallicity, in
agreement with previous works in the literature.  Our NLTE analysis of
the barium lines in the same extremely metal-poor star spectra has led
to some {\it slight} shifts of the derived individual LTE [Ba/Fe]
ratios toward the solar ratio, but the main tendency is only slightly
changed.\\

Globally the results may be summarised as follows (see Fig.~\ref{bafe}):

Below the metallicity $\rm[Fe/H] \approx - 3.0$, the mean value of
[Ba/Fe] is negative as found in Fran\c cois et al.  (\cite{FDH07}).

The slope of the NLTE value of [Ba/Fe] vs.  [Fe/H] is a little
shallower than in LTE: the slope of the regression computed in the
interval $\rm -3.6 <[Fe/H]< -2.6$ is about $1.14 \pm 0.12$ in NLTE
rather than $1.40 \pm 0.18$ under LTE assumption (Fig.~\ref{bafe}).
This small variation of the slope was expected by Mashonkina et al.
(\cite{MZG08}).  The slope is compatible with a secondary production
of Ba.

In the region $\rm -3.6 <[Fe/H]< -2.6$, the scatter around the
regression line of the [Ba/Fe] vs.  [Fe/H] plot (Fig.~\ref{bafe}) is a
little smaller (0.44) for the NLTE values than for the LTE values
(0.52), but it remains very large, much larger than observed for the
abundances of other elements (where it is generally smaller than 0.2
dex).  This scatter strongly exceeds what is expected from the
measurement and determination errors, and illustrates the decoupling
between iron and Ba nucleosyntheses.  This point remains valid when
using Mg (rather than Fe) as a reference element, suggesting also a
decoupling between the production of Mg (hydrostatic burning) and Ba
(r-process).

 In the extremely low metallicity domain ($\rm [Fe/H]<-3.5$), the mean
 behaviour of [Ba/Fe] is not clear.  At this very low metallicity, the
 barium abundance could be estimated in only three "normal" stars.
 (CD~--38:245, CS~22172-002 and CS~22885-096).  In these stars [Ba/Fe]
 is around --0.7 suggesting that a "plateau" of [Ba/Fe] is reached.
 But for two other stars, only limits could be measured and for one of
 them it is rather low: $\rm [Ba/Fe]<-1.37$~.

\subsection{A complex trend ?}
At very low metallicity (~$\rm [Fe/H]<-2.5$), the behaviour of [Ba/Fe]
with [Fe/H] (Fig.~\ref{bafe}) can be interpreted in three different
ways:\\
\begin{enumerate}

\item   
Our Fig.  \ref{bafe} can be interpreted, as proposed by Fran\c cois et
al.  (\cite{FDH07}), by a very large scatter of [Ba/Fe] in the region
$\rm -3.2 <[Fe/H]<-2.8$: a factor of about 1000 from star to star, and
(for $\rm [Fe/H]<-3.3$) a plateau close to the value $\rm
[Ba/Fe]=-1.$\\
This interpretation was suggested by Fran\c cois et al.
(\cite{FDH07}) after merging their measurements with those of Honda et
al.  (\cite{HAK04}).  But it seems that there is a systematic
difference between these two sets of measurements: in Fig.~10 of
Fran\c cois et al.  the data points from Honda et al.  seem to be
systematically offsetted in [Fe/H] and [Ba/Fe] compared to the points
of Fran\c cois et al.  (\cite{FDH07}).  The difference is mainly due
to the adoption of a lower microturbulent velocity by Honda et al.
(\cite{HAK04}).  This systematic shift could be responsible, at least
partly, for the very large scatter without any defined trend,
suggested by Fran\c cois et al.  (\cite{FDH07}) in the region $\rm -
3.2 <[Fe/H]<- 2.8$.

\item   
When only our measurements are taken into account (Fig.~\ref{bafe}),
and as a consequence, all the stars in the graph are studied in a very
homogeneous way, it becomes clear that in the interval $\rm
-3.6<[Fe/H]<2.5$, [Ba/Fe] increases with [Fe/H] with a slope close to
$\rm \approx 1.14$ (when the non-LTE effects are taken into account).

A problem is to know what happens at lower metallicity.  For $\rm
[Fe/H]<-3.6$ the decrease of [Ba/Fe] could continue and then 5 stars
would seem Ba-rich with the definition $\rm [Ba/Fe]>
\overline{[Ba/Fe]} + 2 \sigma$~:\\
\begin{itemize}
\item the two carbon rich stars (CS~22892-052 (Sneden et al.
\cite{SMWP96}, and CS~22949-037 Depagne et al.  \cite{DHS02} ), \item
the well known r-rich star CS~31082-001 (see Cayrel et al., \cite
{CHB01}, Hill et al., \cite {HPC02}), 
\item CD-38:245, the most
metal-poor star of the sample.  In this star Fran\c cois et al.
measured an upper limit of the europium abundance: $\rm
[Eu/Fe]<+0.38$.  According to the definition (Barklem et al.,
\cite{BCB05}, Beers and Christlieb, \cite{BC05}) of the r-I class of
"moderately r-rich stars" ($\rm +0.3 <[Eu/Fe]< +1.0$) the upper limit
of the Eu abundance may bring CD-38:245 in this class.  
\item a
turnoff star, CS~22948-093.  In this kind of star it would be
impossible to measure the europium abundance even if it is r-rich with
[Eu/Fe]=1.
\end{itemize}
All these stars are or could be peculiar (two of them are well known
{\it very} r-rich stars).

However if we suppose that in "normal'' stars [Ba/Fe] decreases
continuously in the interval $\rm -4.2<[Fe/H]<-2.5$ with a uniform
scatter (sigma=0.44), then 4 stars (over 38) are outside the two sigma
limits (3 on one side and 1 on the other side).  This is more than
what it is expected for a Gaussian population (only 5\%).
 
\item   
A third interpretation would combine a decrease of [Ba/Fe] only in the
range $\rm -3.3 <[Fe/H]<-2.5$ and then a plateau which would determine
the mean value of the ratio [Ba/Fe] in the early Galaxy: $\rm
[Ba/Fe]\approx -0.7$ from this NLTE analysis.  The value of this
plateau would define the yields of barium (relative to iron) in the
massive primitive supernovae.  If we consider only the stars with $\rm
[Fe/H] < -3.7$, the scatter around [Ba/Fe] = --0.7 is rather small,
but there are 2 stars with upper limits that increase this scatter.
If a plateau exists, it could suggest a production of Ba independent
of the metallicity in the primitive supernovae.
\end{enumerate} 

\noindent The analysis of a larger sample of EMP stars with $\rm
[Fe/H]<-3.3$ would obviously help in the interpretation of the
results.

\section {Conclusion}
 
We present here a homogeneous determination of the abundance of Ba in
a sample of extremely metal-poor stars (turnoff and giant stars)
taking into account the non-LTE effects.

There is a good agreement between the abundances of turn off stars,
unmixed giants (low RGB) and "mixed" giants (stars located in the HR
diagram higher than the Luminosity function bump).  In the giant stars
the abundance of barium does not depend on the deep mixing inside the
star.

The NLTE abundances are free from the approximations made in the LTE
determinations.  The trend of [Ba/Fe] versus [Fe/H] is refined~: less
scatter, and a better defined and shallower slope.  The slope is
compatible with a secondary process.  The general behaviour would be
the same if Mg (rather than Fe) was the reference element.

The scatter of [Ba/Fe], although reduced, is large, and cannot be
explained by the determination errors: it suggests the influence of
several factors.  The contribution of one (or more) additional "r"
process has been invoked in the literature (see Fran\c cois et al.
and references therein \cite{FDH07}, Travaglio et al., \cite{TGA04},
Wasserburg \& Qian, \cite{WQ08}).  Also, an extremely efficient
galactic mixing is excluded.

The behaviour of [Ba/Fe] with [Fe/H] (Fig.~\ref{bafe}) could be
described in two different ways.  For a decreasing metallicity :
\begin{enumerate}
\item
the mean value of [Ba/Fe] decreases linearly with a large (uniform)
scatter and a few outliers
\item
the mean value of [Ba/Fe] decreases linearly in the interval $\rm
-2.5>[Fe/H]<-3.6$, and then reaches a plateau.
\end{enumerate} 

The number of observed stars at very low metallicity is obviously too
low to reach any firm conclusion about a possible plateau, that could
suggest a production of barium at low metallicity independent of the
metallicity.  Surveys detecting extremely metal-poor stars, and the
spectroscopic observations of the detected stars, are required.

\begin{acknowledgements} The authors thank the referee for a very
careful reading of the manuscript and for useful suggestions.  SMA
kindly acknowledges the support and hospitality of the Paris-Meudon
Observatory.  P.B. acknowledges financial support from E   U contract
MEXT-CT-2004-014265(CIFIST).  M. S., R. C., F. S., P. B., V. H., P. F.
acknowledge the support of CNRS (PNG and PNPS).
\end{acknowledgements}

\end{document}